\documentclass[12pt,pra,aps,showpacs,superscriptaddress]{revtex4}
\usepackage{epsfig}

\def\¦{|}

\begin{document}
\author{Li-Bin Fu}
\affiliation{Institute of Applied Physics and Computational
Mathematics, P.O. Box 8009 (28), 100088 Beijing, China}
\affiliation{Max-Planck-Institute for the Physics of Complex
systems, N\"{o}thnitzer Str. 38, 01187 Dresden, Germany}

\author{Jing-Ling Chen}
\affiliation{Department of Physics, Faculty of Science, National
University of Singapore}

\author{Shi-Gang Chen}
\affiliation{Institute of Applied Physics and Computational
Mathematics, P.O. Box 8009 (28), 100088 Beijing, China}

\title{Maximal violation of Clauser-Horne-Shimony-Holt inequality for four-level
systems}

\begin{abstract}
Clauser-Horne-Shimony-Holt inequality for bipartite systems of
$4$-dimension is studied in detail by employing the unbiased
eight-port beam splitters measurements. The uniform formulae for
the maximum and minimum values of this inequality for such
measurements are obtained. Based on these formulae, we show that
an optimal non-maximally entangled state is about $6\%$ more
resistant to noise than the maximally entangled one. We also give
the optimal state and the optimal angles which are important for
experimental realization.
\end{abstract}

\pacs{03.67.-a, 03.65.-w}
\maketitle


\section{Introduction}

Recently, the inequalities \cite{chb3,gisin} have been found that
generalize the CHSH inequality to systems of high dimension, which
give the analytic description of previous numerical results
\cite{chb1}. In this brief report, we study the CHSH inequality
for bipartite systems for $4$-dimension by employing the unbiased
eight-port beam splitters measurements. The uniform formulae of
the maximal and minimal values of this inequality are obtained.
Based on these formulae, we find an optimal non-maximally
entangled state violates the inequality more strongly than the
maximally entangled one, and then is about $6\%$ more resistant to
noise than the maximally entangled one. Similar to what we have
pointed out for three dimensional systems \cite {fu}, we also find
that the left hand of the inequality can not be violated by the
maximally entangled state. However, we find that the left hand of
the inequality can be violated by some non-maximally entangled
states, and the optimal non-maximally entangled state for the left
hand of the inequality is not the optimal one for the right hand.

\section{The inequality}

In this section, let us recall the Bell inequality for four dimension
obtained in Refs. \cite{gisin,fuin}. The scenario of the inequality involves
two parties: Alice, can carry out two possible measurements, $A_{1}$ or $%
A_{2}$, on one of the particles, whereas the other party, Bob, can carry out
two possible measurements, $B_{1}$ or $B_{2}$, on the other one. For the
composed systems of $4$-dimensional parties (or bipartite systems of spin $%
S=3/2$ particles with the relation $d=2S+1),$ each measurement may have $4$
possible outcomes: $A_{1},A_{2},B_{1},B_{2}=0,1,2,3$. The joint
probabilities are denoted by $P(A_{i},B_{j})$, which are required to satisfy
the normalization condition: $\sum_{m,n=0}^{3}P(A_{i}=m,B_{j}=n)=1$ \cite
{Percival}$.$ We define the correlation functions $Q_{ij}$ as follows,
\begin{equation}
Q_{ij}\equiv \frac{1}{S}\sum_{m=0}^{3}%
\sum_{n=0}^{3}f^{ij}(m,n)P(A_{i}=m,B_{j}=n),  \label{qq}
\end{equation}
in which $S=3/2$, and the function $f^{ij}(m,n)$ can be one of the following
forms,
\begin{equation}
f^{ij}(m,n)=S-M(\varepsilon (i-j)(m\pm n),4),  \label{fff}
\end{equation}
where $\varepsilon (x)$ is the sign function: $\varepsilon (x)=\left\{
\begin{array}{cc}
1 & x\geq 0 \\
-1 & x<0
\end{array}
\right. ,$ and $M(x,4)$ is defined as follows: $M(x,4)=(x\mbox{
mod 4})$ and $0\leq M(x,4)\leq 3$. Then one can consider the following Bell
expression:
\begin{equation}
I_{4}=Q_{11}+Q_{12}-Q_{21}+Q_{22}.  \label{bd}
\end{equation}

Especially, if taking $f^{ij}(m,n)=S-M\left( \varepsilon (i-j)(m-n),4\right)
$, one can prove that the above expression is just the Bell expression
presented in Ref. \cite{gisin}$.$

The authors of \cite{gisin,fuin} proved that the maximum value of the above
Bell expressions is $2$ and minimum value of it is $-10/3$ for local
variable theories. Then we get the following Bell inequality:
\begin{equation}
-\frac{10}{3}\leq I_{4}\leq 2.  \label{ineq}
\end{equation}

However, the inequality will be violated for some entangled states by
quantum predictions. For the maximally entangled state of two $4$%
-dimensional systems $\psi =\frac{1}{2}\sum_{j=0}^{3}\left| l\right\rangle
_{A}\left| l\right\rangle _{B},$ the authors of \cite{gisin,fuin} obtained
the strongest violation of Bell expression $I_{4}$ for such a state, $%
I_{4}(QM)=\frac{2}{3}\left( \sqrt{2}+\sqrt{10-\sqrt{2}}\right) \approx
2.89624.$ In the presence of uncolored noise the quantum state
\begin{equation}
\rho =(1-F)\left| \psi \right\rangle \left\langle \psi \right| +F\frac{I}{16}%
.  \label{nose}
\end{equation}
Following \cite{chb1}, we define the threshold noise admixture $F_{thr}$
(the minimal noise admixture fraction for $\left| \psi \right\rangle $) $%
F_{thr}=1-2/I_{4}(QM),$ then\ we get $F_{thr}\approx 0.30945.$ This result
equals to the numerical results of\ \cite{chb1}.

\section{The maximal violation}

Here, as in the previous works, we restricted our analysis to unbiased
eight-port beam splitters \cite{trit,chb7}, more specifically to Bell
multi-port beam splitters \cite{chb1}. Unbiased Bell $2d$-port beam
splitters have the property that a photon entering into any single port (out
of the $d$) has equal chances to exit from any output port. In additional,
for Bell $2d$-port beam splitters the elements of their unitary transition
matrix, ${\bf U}^{d},$ are solely powers of the $d$-th root of unity $\gamma
_{d}=\exp (i2\pi /d),$ namely, ${\bf U}_{ij}^{d}=\frac{1}{\sqrt{d}}\gamma
_{d}^{(i-1)(j-1)}.$

Let us now imagine spatially separated Alice and Bob who perform the Bell
type experiment via the eight-port beam splitters on the state
\begin{equation}
\left| \phi \right\rangle =\frac{1}{2}\sum\limits_{i=0}^{3}a_{i}\left|
i\right\rangle _{A}\left| i\right\rangle _{B},  \label{psi}
\end{equation}
with real coefficients $a_{i},$ where, e.g., $\left| i\right\rangle _{A}$
describes a photon in mode $i$ propagating to Alice. One has $%
_{x}\left\langle i|i^{\prime }\right\rangle _{x}=\delta _{ii^{\prime }},$
with $x=A,B.$ The overall unitary transformation performed by such a device
is given by
\begin{equation}
U_{ij}^{4}=\frac{1}{2}\gamma _{4}^{ij}e^{i\varphi _{j}},\;\;i,j=0,1,2,3
\label{u}
\end{equation}
where $\gamma _{4}=e^{i\pi /2}$ and $j$ denotes an input beam to the device,
and $i$ an output one; $\varphi _{j}$ are the four phases that can be set by
the local observer, denoted as $\vec{\varphi}=(\varphi _{0},\varphi
_{1},\varphi _{2},\varphi _{3})$ . The transformations at Alice's side are
denoted as $\vec{\varphi}^{A}=(\varphi _{0}^{A},\varphi _{1}^{A},\varphi
_{2}^{A},\varphi _{3}^{A}),$ and $\vec{\varphi}^{B}=(\varphi
_{0}^{B},\varphi _{1}^{B},\varphi _{2}^{B},\varphi _{3}^{B})$ for Bob's
side. In this way the local observable is defined. The quantum prediction
for the joint probability $P^{QM}(A_{i}=m,B_{j}=n)$ to detect a photon at
the $m$-th output of the multiport $A$ and another one at the $n$-th output
of the multiport $B$ calculated for the state (\ref{psi}) is given by
\begin{equation}
P^{QM}(A_{i}=m,B_{j}=n)=\sum\limits_{k=0}^{3}\sum\limits_{l=0}^{3}a_{k}a_{l}%
\gamma _{4}^{(k-l)(m+n)}e^{i(\varphi _{k}^{A_{i}}+\varphi
_{k}^{B_{j}}-\varphi _{l}^{A_{i}}-\varphi _{l}^{B_{j}})},  \label{pqm}
\end{equation}

For convenience, we use the definition of the correlation functions $Q_{ij}$
with the function $f^{ij}=S-M(\varepsilon (i-j)(m+n),d).$ By substituting (%
\ref{pqm}) into (\ref{bd}), after some elaborate, we can obtain
\begin{equation}
I_{4}=\sum_{k<l}a_{k}a_{l}T_{kl},\text{ }(k=0,1,2;l=1,2,3),  \label{ss}
\end{equation}
where $T_{kl}$ are six continuous functions of sixteen angles $\vec{\varphi}%
^{A_{i}}$ and $\vec{\varphi}^{B_{j}}$ $(i,j=1,2).$ Let us define $\varphi
_{ab}^{ij}=\varphi _{a}^{A_{i}}-\varphi _{b}^{A_{i}}+\varphi
_{a}^{B_{j}}-\varphi _{b}^{B_{j}},$ then we can list these six functions as
the follows:
\begin{eqnarray}
T_{01} &=&\frac{1}{6}\left\{ \left[ \cos (\varphi _{01}^{11})-\cos (\varphi
_{01}^{21})-\cos (\varphi _{01}^{12})+\cos (\varphi _{01}^{22})\right]
\right. -  \nonumber \\
&&\left. \left[ \sin (\varphi _{01}^{11})+\sin (\varphi _{01}^{21})+\sin
(\varphi _{01}^{12})+\sin (\varphi _{01}^{22})\right] \right\} ,  \label{t01}
\end{eqnarray}
\begin{equation}
T_{02}=-\frac{1}{6}\left[ \cos (\varphi _{02}^{11})-\cos (\varphi
_{02}^{21})+\cos (\varphi _{02}^{12})+\cos (\varphi _{02}^{22})\right] ,
\label{t02}
\end{equation}
\begin{eqnarray}
T_{03} &=&\frac{1}{6}\left\{ \left[ \cos (\varphi _{03}^{11})-\cos (\varphi
_{03}^{21})-\cos (\varphi _{03}^{12})+\cos (\varphi _{03}^{22})\right]
\right. +  \nonumber \\
&&\left. \left[ \sin (\varphi _{03}^{11})+\sin (\varphi _{03}^{21})+\sin
(\varphi _{03}^{12})+\sin (\varphi _{03}^{22})\right] \right\} ,  \label{03}
\end{eqnarray}
\begin{eqnarray}
T_{12} &=&\frac{1}{6}\left\{ \left[ \cos (\varphi _{12}^{11})-\cos (\varphi
_{12}^{21})-\cos (\varphi _{12}^{12})+\cos (\varphi _{12}^{22})\right]
\right. -  \nonumber \\
&&\left. \left[ \sin (\varphi _{12}^{11})-\sin (\varphi _{12}^{21})+\sin
(\varphi _{12}^{12})+\sin (\varphi _{12}^{22})\right] \right\} ,  \label{04}
\end{eqnarray}
\begin{equation}
T_{13}=-\frac{1}{6}\left[ \cos (\varphi _{13}^{11})-\cos (\varphi
_{13}^{21})+\cos (\varphi _{13}^{12})+\cos (\varphi _{13}^{22})\right] ,
\label{13}
\end{equation}
and
\begin{eqnarray}
T_{23} &=&\frac{1}{6}\left\{ \left[ \cos (\varphi _{23}^{11})-\cos (\varphi
_{23}^{21})-\cos (\varphi _{23}^{12})+\cos (\varphi _{23}^{22})\right]
\right. -  \nonumber \\
&&\left. \left[ \sin (\varphi _{23}^{11})-\sin (\varphi _{23}^{21})+\sin
(\varphi _{23}^{12})+\sin (\varphi _{23}^{22})\right] \right\} ,  \label{23}
\end{eqnarray}

We can know that $|T_{kl}|\leq \frac{\sqrt{10-\sqrt{2}}\left( 2+3\sqrt{2}%
\right) }{21}.$ However, they are strongly correlated, so $T_{kl}$
can not reach their maximum value at the same time. As has been
proven in Ref. \cite {fu}, the maximum (minimum) values of $I_{4}$
can only be found on the vertices of the polyhedron formed by
$T_{kl}$. There are three sets of the
vertices of the polyhedron. By denoting $\Gamma _{1}=\frac{\sqrt{10-\sqrt{2}}%
\left( 2+3\sqrt{2}\right) }{21}\approx 0.87104,$ $\Gamma _{2}=\frac{\sqrt{2}%
}{3}\approx 0.4714,$ and $\Gamma _{3}=\frac{\sqrt{10-\sqrt{2}}\left( 4-\sqrt{%
2}\right) }{21}\approx 0.3608,$ we list them in the following three tables.

In these tables, we have used the stipulations, e.g., $T_{[ab]}=T_{ab}$ for $%
a<b$ or $T_{[ab]}=T_{ba}$ for $b<a.$ As has been proven in Ref. \cite{fu},
for each vertices of the polyhedron one can obtain an extreme value of the $%
I_{4},$ and the maximum (or minimum) value can be obtained by comparing the
values among them.

Assuming $%
\{A_{i},\ (i=0,1,2,3)\}=$ $\{|a_{0}|,|a_{1}|,|a_{2}|,|a_{3}|\}$ $,$ where $%
``="$ means the equality of two sets, and $A_{i}$ are in decreasing order,
i.e. $A_{0}\geq A_{1}\geq A_{2}\geq A_{3}.$ For convenience, we denote $%
\Gamma _{1}=\frac{\sqrt{10-\sqrt{2}}\left( 2+3\sqrt{2}\right) }{21},$ $%
\Gamma _{2}=\frac{\sqrt{2}}{3},$ and $\Gamma _{3}=\frac{\sqrt{10-\sqrt{2}}%
\left( 4-\sqrt{2}\right) }{21}.$ Let us define
\begin{equation}
B_{1}(\left| \phi \right\rangle )=(A_{0}A_{1})\Gamma
_{1}+(A_{0}A_{2}+A_{1}A_{3})\Gamma
_{2}+(A_{0}A_{3}+A_{1}A_{2}+A_{2}A_{3})\Gamma _{3},  \label{ss1}
\end{equation}
and
\begin{equation}
B_{2}(\left| \phi \right\rangle )=(A_{0}A_{1})\Gamma
_{3}+(A_{0}A_{2}+A_{1}A_{3})\Gamma
_{2}+(A_{0}A_{3}+A_{1}A_{2}-A_{2}A_{3})\Gamma _{1}.  \label{ss2}
\end{equation}
Then, the maximum value of $I_{4}$ for $\left| \phi \right\rangle
$ must be
\begin{equation}
I_{4}^{\max }(\left| \phi \right\rangle )=Max(B_{1}(\left| \phi
\right\rangle ),B_{2}(\left| \phi \right\rangle )).  \label{ssmax}
\end{equation}

From (\ref{ssmax}), one can immediately get that for the maximally
entangled states, i.e., $|a_{i}|=1$ $(i=0,1,2,3),$ the maximum value of $%
I_{4}$ for such states are
\begin{equation}
I_{4}^{\max }=\Gamma _{1}+2\Gamma _{2}+3\Gamma _{3}=\frac{2}{3}\left( \sqrt{2%
}+\sqrt{10-\sqrt{2}}\right) .
\end{equation}
This is just the result obtained in Ref. \cite{gisin,acin,fuin}, and gives $%
F_{thr}\approx 0.30945$ which equals to the numerical results of\ \cite{chb1}%
.

Consider $a_{i}$ as variables, we can obtain the maximal value of $%
I_{4}^{\max },$ denoted as $\bar{I}_{\max }.$ The value of Eq. (\ref{ssmax})
should be maximum for $A_{0}=A_{1}$ and $A_{2}=A_{3}.$ One can easily find
that $B_{1}(\left| \phi \right\rangle )>B_{2}(\left| \phi \right\rangle )$
for any state$.$ For this case, By calculating the extreme value of $%
B_{1}(\left| \phi \right\rangle )$ with $A_{0}=A_{1}$ and $A_{2}=A_{3},$
after some elaboration, we get
\begin{equation}
\bar{I}_{\max }=\bar{A}_{+}^{2}\Gamma _{1}+2\bar{A}_{+}\bar{A}_{-}(\Gamma
_{2}+\Gamma _{3})+\bar{A}_{-}^{2}\Gamma _{3},  \label{smax1}
\end{equation}
where
\[
\bar{A}_{\pm }=\sqrt{1\pm \sqrt{\frac{1}{791}\left( 357+7\sqrt{2}-20\sqrt{10-%
\sqrt{2}}-58\sqrt{2(10-\sqrt{2})}\right) }}
\]
with $\bar{A}_{+}=A_{0}=A_{1}$ and $\bar{A}_{-}=A_{2}=A_{3}.$ We then have $%
\bar{S}_{\max }\approx 2.9727,\;$and the threshold amount of noise is about $%
F_{thr}\approx 0.3272,$ which was also obtained in Ref. \cite{acin} by
calculating the maximum eigenvalue of the Bell operator \cite{acin25}. One
can see this optimal non-maximally entangled state is about $6\%$ more
resistant to noise than the maximally entangled one. One can check the above
results with the following optimal angles for $a_{i}>0$ $(i=0,1,2,3):$%
\begin{equation}
\vec{\varphi}^{A_{1}}=(0,\frac{\pi }{6},-\pi ,\frac{4\pi }{9}),\;\;\vec{%
\varphi}^{A_{2}}=(0,-\frac{5\pi }{9},\frac{5\pi }{9},-\frac{\pi }{3});
\label{ang1}
\end{equation}
and
\begin{equation}
\vec{\varphi}^{B_{1}}=(0,-\frac{\pi }{2},\frac{13\pi }{18},-\frac{11\pi }{18}%
),\;\;\vec{\varphi}^{B_{2}}=(0,\frac{7\pi }{36},-\frac{27\pi }{36},-\frac{%
7\pi }{18}).  \label{ang2}
\end{equation}

On the other hand, we can also calculated the minimum value of $I_{4}.$ Let
us define
\begin{equation}
S_{1}(\left| \phi \right\rangle )=-(A_{0}A_{1}+A_{0}A_{3}+A_{1}A_{2})\Gamma
_{1}-(A_{0}A_{2}+A_{1}A_{3})\Gamma _{2}+(A_{2}A_{3})\Gamma _{3},  \label{ma1}
\end{equation}
and
\begin{equation}
S_{2}(\left| \phi \right\rangle
)=-2(A_{0}A_{1}+A_{0}A_{3}+A_{1}A_{2}+A_{2}A_{3})/3-(A_{0}A_{2}+A_{1}A_{3})/3,
\label{ma2}
\end{equation}
The minimum value of $I_{4}$ for $\left| \phi \right\rangle $ should be
\begin{equation}
I_{4}^{\min }(\left| \phi \right\rangle )=Min(S_{1}(\left| \phi
\right\rangle ),S_{2}(\left| \phi \right\rangle )).  \label{mm1}
\end{equation}
Then, for the maximally entangled state, the minimum value of $I_{4}$ is
\begin{equation}
I_{4}^{\min }=-\frac{10}{3}.
\end{equation}
One sees that the maximally entangled state does not violate the left hand
of the inequality (\ref{bd}). However, for a non-maximally entangled state
with $A_{0}=A_{1}=K_{+}$ and $A_{2}=A_{3}=K_{-}$ where
\begin{equation}
K_{\pm }=\sqrt{1\pm \sqrt{\frac{1}{791}\left( 357-7\sqrt{2}-80\sqrt{10-\sqrt{%
2}}+6\sqrt{2(10-\sqrt{2})}\right) },}
\end{equation}
we find the minimum value of $I_{4}^{\min },$ denoted as $\bar{I}_{\min }$
\begin{equation}
\bar{I}_{\min }=-K_{+}^{2}\Gamma _{1}-2K_{+}K_{-}(\Gamma _{1}+\Gamma
_{2})+K_{-}^{2}\Gamma _{3}\approx -3.46424.  \label{mmmin}
\end{equation}
Obviously, such states violate the left hand of the inequality.

\section{Discussion}

In summary of the present paper, we study the CHSH inequality for
$d=4$ in details on the Bell type experiment via the eight-port
beam splitters which is realizable for nowadays technique. We give
the analytic formulae of the maximum and minimum values of this
inequality for such an experimental consideration. The maximal
violations we obtained are the same as Refs. \cite{acin}. We also
give the optimal state and the optimal angles which are important
for experimental realization.

It is well-known that for bipartite systems of $2$-dimension, the
CHSH inequality is symmetry. For any entangled state the
inequality is violated symmetrically, and will be maximally
violated by the maximally entangled states. However, for the
higher dimensional systems, namely $d$-dimensional systems
($d>2$), the inequality is asymmetry (see in Ref. \cite
{gisin,fuin}),
\begin{equation}
\frac{-2(d+1)}{(d-1)}\leq I_{d}\leq 2.  \label{ined}
\end{equation}
The authors of Ref. \cite{gisin,fuin} studied the violation of the
right hand of the above inequality for maximally entangled states
and reproduced the results of previous numerical works
\cite{chb1}. For $d=3,$ one can immediately obtain that the left
hand of the above inequality is $-4,$ which
can never be violated by any state \cite{gisin,fuin}. In Ref. \cite{fu}%
, the authors shown that the minimal values of $I_{3}$\ for
maximally entangled states is just $-4.\ $They also found that an
optimal non-maximally entangled state violates the inequality more
strongly than the maximally entangled one. For $d=4$, we also find
that the left hand of the inequality can be violated by some
non-maximally entangled states, and the optimal non-maximally
entangled state for the left hand of the inequality is not the
optimal one for the right hand.

In fact, the relations (\ref{ined}) and (\ref{bd}) define two
inequalities, namely, the right ones and the left ones. The right
inequalities are optimal and tight but the left ones are not
tight. The asymmetry of the CHSH inequalities is due to the
asymmetry of Hilbert space for higher dimensional systems
\cite{fuadl}. For the systems of $2$-dimension, the Hilbert space
is symmetry, so the CHSH inequalities are symmetry as well. But
for the higher dimensional systems, the Hilbert space is
asymmetry, so the inequalities which are optimal for the right
hand will be not optimal for the left. In other word, we can not
find a inequality for higher dimensional systems which is optimal
for the both sides.

\section{Acknowledgment}

This work was supported by the 973 Project of China and the
Alexander von Humboldt Foundation..

\begin{table}[th]
\begin{ruledtabular}
\begin{tabular}{||c|c|c|c|c|c|c||}
$\left\{ T_{ij}\right\} $ & $T_{[ab]}$~~~~~~~~~~~ & $T_{[ac]}$~~~~~~~~~~~ & $%
T_{[ad]}$~~~~~~~~~~~ & $T_{[bc]}$~~~~~~~~~~~ & $T_{[bd]}$~~~~~~~~~~~ & $%
T_{[cd]}$~~~~~~~~~~~ \\ \hline
& $\Gamma _{1}$~~~~~~~~~~~ & $\Gamma _{2}$~~~~~~~~~~~ & $\Gamma _{3}$%
~~~~~~~~~~~ & $\Gamma _{3}$~~~~~~~~~~~ & $\Gamma _{2}$~~~~~~~~~~~
& $\Gamma _{3}$~~~~~~~~~~~ \\ \hline
& $-\Gamma _{1}$~~~~~~~~~~~~ & $-\Gamma _{2}$~~~~~~~~~~~~ & $-\Gamma _{3}$%
~~~~~~~~~~~~ & $\Gamma _{3}$~~~~~~~~~~~ & $\Gamma _{2}$~~~~~~~~~~~
& $\Gamma _{3}$~~~~~~~~~~~ \\ \hline
& $-\Gamma _{1}$~~~~~~~~~~~~ & $\Gamma _{2}$~~~~~~~~~~~ & $\Gamma _{3}$%
~~~~~~~~~~~ & $-\Gamma _{3}$~~~~~~~~~~~~ & $-\Gamma _{2}$~~~~~~~~~~~~ & $%
\Gamma _{3}$~~~~~~~~~~~ \\ \hline
& $\Gamma _{1}$~~~~~~~~~~~ & $-\Gamma _{2}$~~~~~~~~~~~~ & $\Gamma _{3}$%
~~~~~~~~~~~ & $-\Gamma _{3}$~~~~~~~~~~~~ & $\Gamma _{2}$~~~~~~~~~~~ & $%
-\Gamma _{3}$~~~~~~~~~~~~ \\ \hline
& $\Gamma _{1}$~~~~~~~~~~~ & $\Gamma _{2}$~~~~~~~~~~~ & $-\Gamma _{3}$%
~~~~~~~~~~~~ & $\Gamma _{3}$~~~~~~~~~~~ & $-\Gamma _{2}$~~~~~~~~~~~~ & $%
-\Gamma _{3}$~~~~~~~~~~~~ \\ \hline
& $\Gamma _{1}$~~~~~~~~~~~ & $-\Gamma _{2}$~~~~~~~~~~~~ & $-\Gamma _{3}$%
~~~~~~~~~~~~ & $-\Gamma _{3}$~~~~~~~~~~~~ & $-\Gamma _{2}$~~~~~~~~~~~~ & $%
\Gamma _{3}$~~~~~~~~~~~ \\ \hline
& $-\Gamma _{1}$~~~~~~~~~~~~ & $\Gamma _{2}$~~~~~~~~~~~ & $-\Gamma _{3}$%
~~~~~~~~~~~~ & $-\Gamma _{3}$~~~~~~~~~~~~ & $\Gamma _{2}$~~~~~~~~~~~ & $%
-\Gamma _{3}$~~~~~~~~~~~~ \\ \hline
& $-\Gamma _{1}$~~~~~~~~~~~~ & $-\Gamma _{2}$~~~~~~~~~~~~ & $\Gamma _{3}$%
~~~~~~~~~~~ & $\Gamma _{3}$~~~~~~~~~~~ & $-\Gamma _{2}$~~~~~~~~~~~ & $%
-\Gamma _{3}$~~~~~~~~~~~~
\end{tabular}
\medskip
\caption{The first set of vertices, in which $i\neq j$. }
\label{summary}
\end{ruledtabular}
\end{table}

\begin{table}[th]
\begin{ruledtabular}
\begin{tabular}{||c|c|c|c|c|c|c||}
$\left\{ T_{ij}\right\} $ & $T_{[ab]}$~~~~~~~~~~~ & $T_{[ac]}$~~~~~~~~~~~ & $%
T_{[ad]}$~~~~~~~~~~~ & $T_{[bc]}$~~~~~~~~~~~ & $T_{[bd]}$~~~~~~~~~~~ & $%
T_{[cd]}$~~~~~~~~~~~ \\ \hline
& $-\Gamma _{1}$~~~~~~~~~~~~ & $-\Gamma _{2}$~~~~~~~~~~~~ & $-\Gamma _{1}$%
~~~~~~~~~~~~ & $-\Gamma _{1}$~~~~~~~~~~~~ & $-\Gamma _{2}$~~~~~~~~~~~~ & $%
\Gamma _{3}$~~~~~~~~~~~ \\ \hline & $\Gamma _{1}~~~~~~~~~~~$ &
$\Gamma _{2}~~~~~~~~~~~~$ & $\Gamma _{1}~~~~~~~~~~~$ & $-\Gamma
_{1}$~~~~~~~~~~~~ & $-\Gamma _{2}$~~~~~~~~~~~~ & $\Gamma
_{3}$~~~~~~~~~~~ \\ \hline & $\Gamma _{1}~~~~~~~~~~~$ & $-\Gamma
_{2}~~~~~~~~~~~~$ & $-\Gamma
_{1}~~~~~~~~~~~~$ & $\Gamma _{1}$~~~~~~~~~~~ & $\Gamma _{2}$~~~~~~~~~~~ & $%
\Gamma _{3}$~~~~~~~~~~~ \\ \hline & $-\Gamma _{1}~~~~~~~~~~~~$ &
$\Gamma _{2}~~~~~~~~~~~$ & $-\Gamma _{1}~~~~~~~~~~~~$ & $-\Gamma
_{1}$~~~~~~~~~~~~ & $-\Gamma _{2}$~~~~~~~~~~~~ & $-\Gamma
_{3}$~~~~~~~~~~~~ \\ \hline & $-\Gamma _{1}~~~~~~~~~~~~$ &
$-\Gamma _{2}~~~~~~~~~~~~$ & $\Gamma
_{1}~~~~~~~~~~~$ & $-\Gamma _{1}$~~~~~~~~~~~~ & $\Gamma _{2}$~~~~~~~~~~~ & $%
-\Gamma _{3}$~~~~~~~~~~~~ \\ \hline & $-\Gamma _{1}~~~~~~~~~~~~$ &
$\Gamma _{2}~~~~~~~~~~~$ & $\Gamma
_{1}~~~~~~~~~~~$ & $\Gamma _{1}$~~~~~~~~~~~ & $\Gamma _{2}$~~~~~~~~~~~ & $%
\Gamma _{3}$~~~~~~~~~~~ \\ \hline & $\Gamma _{1}~~~~~~~~~~~$ &
$-\Gamma _{2}~~~~~~~~~~~~$ & $\Gamma
_{1}~~~~~~~~~~~$ & $\Gamma _{1}$~~~~~~~~~~~ & $-\Gamma _{2}$~~~~~~~~~~~~ & $%
-\Gamma _{3}$~~~~~~~~~~~~ \\ \hline & $\Gamma _{1}~~~~~~~~~~~$ &
$\Gamma _{2}~~~~~~~~~~~$ & $-\Gamma
_{1}~~~~~~~~~~~~$ & $-\Gamma _{1}$~~~~~~~~~~~~ & $\Gamma _{2}$~~~~~~~~~~~ & $%
-\Gamma _{3}~$~~~~~~~~~~~
\end{tabular}
\medskip
\caption{The second set of vertices, in which $i\neq j$.}
\label{summary1}
\end{ruledtabular}
\end{table}

\begin{table}[th]
\begin{ruledtabular}
\begin{tabular}{||c|c|c|c|c|c|c||}
$\left\{ T_{ij}\right\} $ & $T_{[ab]}$~~~~~~~~~~~ & $T_{[ac]}$~~~~~~~~~~~ & $%
T_{[ad]}$~~~~~~~~~~~ & $T_{[bc]}$~~~~~~~~~~~ & $T_{[bd]}$~~~~~~~~~~~ & $%
T_{[cd]}$~~~~~~~~~~~ \\ \hline
& $-\frac{2}{3}~$~~~~~~~~~~~ & $-\frac{1}{3}~$~~~~~~~~~~~~ & $-\frac{2}{3}~$%
~~~~~~~~~~~~ & $-\frac{2}{3}~$~~~~~~~~~~~~ & $-\frac{1}{3}~$~~~~~~~~~~~~ & $-%
\frac{2}{3}~$~~~~~~~~~~~~ \\ \hline
& $\frac{2}{3}~$~~~~~~~~~~ & $\frac{1}{3}~$~~~~~~~~~~~ & $\frac{2}{3}~$%
~~~~~~~~~~~ & $-\frac{2}{3}~$~~~~~~~~~~~~ & $-\frac{1}{3}~$~~~~~~~~~~~~ & $-%
\frac{2}{3}~$~~~~~~~~~~~~ \\ \hline
& $\frac{2}{3}~$~~~~~~~~~~ & $-\frac{1}{3}~$~~~~~~~~~~~~ & $-\frac{2}{3}~$%
~~~~~~~~~~~~ & $\frac{2}{3}~$~~~~~~~~~~~ & $\frac{1}{3}~$~~~~~~~~~~~ & $-%
\frac{2}{3}~$~~~~~~~~~~~ \\ \hline
& $-\frac{2}{3}~$~~~~~~~~~~~ & $\frac{1}{3}~$~~~~~~~~~~~ & $-\frac{2}{3}~$%
~~~~~~~~~~~~ & $\frac{2}{3}~$~~~~~~~~~~~ & $-\frac{1}{3}~$~~~~~~~~~~~~ & $%
\frac{2}{3}~$~~~~~~~~~~~ \\ \hline
& $-\frac{2}{3}~$~~~~~~~~~~~ & $-\frac{1}{3}~$~~~~~~~~~~~~ & $\frac{2}{3}~$%
~~~~~~~~~~~ & $-\frac{2}{3}~$~~~~~~~~~~~~ & $\frac{1}{3}~$~~~~~~~~~~~ & $%
\frac{2}{3}~$~~~~~~~~~~~ \\ \hline
& $-\frac{2}{3}~$~~~~~~~~~~~ & $\frac{1}{3}~$~~~~~~~~~~~ & $\frac{2}{3}~$%
~~~~~~~~~~~ & $\frac{2}{3}~$~~~~~~~~~~~ & $\frac{1}{3}~$~~~~~~~~~~~ & $-%
\frac{2}{3}~$~~~~~~~~~~~ \\ \hline
& $\frac{2}{3}~$~~~~~~~~~~ & $-\frac{1}{3}~$~~~~~~~~~~~~ & $\frac{2}{3}~$%
~~~~~~~~~~~ & $\frac{2}{3}~$~~~~~~~~~~~ & $-\frac{1}{3}~$~~~~~~~~~~~~ & $%
\frac{2}{3}~$~~~~~~~~~~~ \\ \hline
& $\frac{2}{3}~$~~~~~~~~~~ & $\frac{1}{3}~$~~~~~~~~~~~ & $-\frac{2}{3}~$%
~~~~~~~~~~~~ & $-\frac{2}{3}~$~~~~~~~~~~~~ & $\frac{1}{3}~$~~~~~~~~~~~ & $%
\frac{2}{3}~$~~~~~~~~~~~
\end{tabular}
\medskip
\caption{The third set of vertices, in which $i\neq j$}
\label{summary2}
\end{ruledtabular}
\end{table}

\end{document}